\documentclass[sigconf]{acmart}

\usepackage{algpseudocode} 
\usepackage{algorithm}

\usepackage{subfigure}
\usepackage{graphicx}

\usepackage{pgfplots}
\usepackage{pgfplotstable}
\pgfplotsset{compat=newest}
 
\acmYear{2021}\copyrightyear{2021}
\setcopyright{rightsretained}
\acmConference[IoTDI '21]{International Conference on Internet-of-Things Design and Implementation}{May 18--21, 2021}{Charlottesvle, VA, USA}
\acmBooktitle{International Conference on Internet-of-Things Design and Implementation (IoTDI '21), May 18--21, 2021, Charlottesvle, VA, USA}
\acmPrice{15.00}
\acmDOI{10.1145/3450268.3453510}
\acmISBN{978-1-4503-8354-7/21/05}

\begin{document}

\title{Optimal Distributed Bandwidth Allocation in NB-IoT Networks}

\author{Hongde Wu, Zhengyong Chen, Noel E. O'Connor and Mingming Liu}
\affiliation{%
	\institution{School of Electronic Engineering, Dublin City University, Dublin 9, Ireland}}
\email{{hongde.wu7, zhengyong.chen5}@mail.dcu.ie} \email{{noel.oconnor, mingming.liu}@dcu.ie}

\begin{abstract}
In this paper, we investigate a key problem of Narrowband-Internet of Things (NB-IoT) in the context of 5G with Mobile Edge Computing (MEC). We address the challenge that  IoT devices may have different priorities when demanding bandwidth for data transmission in specific applications and services. Due to the scarcity of bandwidth in an MEC enabled IoT network,  our objective is to optimize bandwidth allocation for a group of NB-IoT devices in a way that the group can work collaboratively to maximize their overall utility. To this end, we design an optimal distributed algorithm and use simulations to demonstrate its efficacy to effectively manage various IoT data streams in a fully distributed framework. 
\end{abstract}





\maketitle

\section{Introduction}

In recent years, Internet of Things (IoT) has gradually become an increasingly important computing paradigm, given its potential to provide seamless network connectivity between people, machines and things.  In a typical network setup, IoT data can be collected by sensors, transmitted to a gateway, processed and stored on the cloud, and shared for various application purposes. 

In the context of 5G, data from heterogeneous IoT devices can be transmitted to an edge server within the range of radio access networks, where it can be pocessed i.e. Mobile Edge Computing (MEC). By doing this, key functions from the cloud can be deployed closer to the IoT network, providing the low latency and highly reliable network services needed for various computation and storage tasks. On the other hand, ensuring low power consumption and massive connections of devices is also important in IoT applications. In this context, the Third Generation Partnership Project (3GPP) introduced a wireless access technology namely Narrowband-Internet of Things (NB-IoT), designed to utilize limited licensed spectrum in existing mobile networks to handle a limited amount of bidirectional data transmission between devices and base stations. Technically,  NB-IoT has a gain of 20dB compared to existing networks in the same radio band, which is equivalent to a hundred-fold increase in the capacity of the coverage area. Clearly, this feature is a significant enabler for Machine-to-Machine (M2M) communication capabilities for IoT in 5G.

An interesting research problem for NB-IoT is the bandwidth allocation problem. For instance, in \cite{beshley2018method}, a centralized controller was used to allocate spectrum resources; in \cite{ejaz2017multiband}, the authors proposed multi-wideband spectrum methods to allocate bandwidth in 5G networks for IoT applications with different Quality of Service (QoS) requirements; and in \cite{halabian2019distributed} convex optimization was used to maximize the overall utility values of network slices. In this paper, we propose an optimal bandwidth management system for IoT edge devices in a distributed architecture.  Our objective is to maximize the overall utility of the group of IoT devices given the available bandwidth in the network. The utility refers to how an IoT device can practically benefit from a given bandwidth. Specifically, we assume that different IoT devices may have different priorities in data transmission, which can lead to different utility values for the applications. In the following sections, we will show  how the problem  can be formulated as a concave optimization problem with constraints. We address the problem in a distributed manner without comprising users' privacy. 



\section{System Model}

We  consider a typical IoT scenario where a number of $N$ devices are connected to the MEC server. We model the utility function of the $i^{th}$ device, $U_i(x_i)$, with respect to its bandwidth $x_i$, as follows:

\begin{equation} \label{uf}
U_i(x_i) = \omega_i \log ( x_i \log_2 (1 + \frac{S}{R} ) + 1 ) - p * x_i^{2}
\end{equation}

\noindent where  $x_i \log_2 (1 + \frac{S}{R} )$ models the transmission rate of  device $i$ given its bandwidth $x_i$ and the signal-noise-ratio $\frac{S}{R}$ as per  Shannon's theorem. As a result, the $\omega_i \log (  x_i \log_2 (1 + \frac{S}{R} ) + 1 )$ term models the utility that the $i^{th}$ device can obtain with respect to the bandwidth $x_i$, in which $\omega_i$ is a parameter that reflects the level of QoS that the $i^{th}$ device expects, and the $``+1"$ term models that the utility expected from the device is 0 when the transmission rate is 0. Finally, we also consider a simple pricing model when a group of devices share the total bandwidth $B$ available in the network. For this purpose we assume that the cost of data transmission of device $i$ is proportional to the square of bandwidth, i.e. $p * x_i^{2}$, where $p$ is a positive pricing coefficient. In this regard, $U_i(x_i)$ can be seen as the net utility of the device $i$ with respect to $x_i$. Since our purpose is to illustrate the concept of the proposed framework and the distributed algorithm, the units of the parameters
are ignored in simulation and discussion.

\section{Problem Formulation}

Our objective in this paper can be formulated as follows. 

\begin{equation} \label{eq}
	\begin{gathered}
		\underset{x_1, x_2, \ldots, x_{N}}{\max} \quad
		\sum\limits_{i = 1}^{N} U_{i}\left(x_i \right), \quad 
		{\text{s.t.}} ~
		\sum\limits_{i = 1}^{N} x_i = \sum\limits_{i = 1}^{N} d_i \leq B ,  \\
	\end{gathered}
\end{equation}

\noindent where $d_i$ denotes the desired bandwidth that the $i^{th}$ device wishes to obtain, and it is a local variable that can only be accessed by  device $i$. We note that $d_i$ is a target or expected value that the $i$'th device wishes to achieve for $x_i$, and there is no guarantee that $x_i = d_i$ when an optimal solution is found as our optimization problem is to ensure $\sum\limits_{i = 1}^{N} x_i = \sum\limits_{i = 1}^{N} d_i$. The importance of introducing $d_i$ is that each local device can only get access to $d_i$ for computing its bandwidth, and this enables better privacy protection  for both IoT devices and the MEC server, i.e. no direct access to $B$ or $d_j, \forall j \neq i$.

\section{Algorithm and Optimization}

We now present the proposed optimal distributed algorithm for solving the problem given in Equation \eqref{eq} inspired by the work reported in \cite{yi2016initialization}.

\begin{algorithm}[htbp]
	\caption{Optimal Distributed Algorithm for MEC NB-IoT}
\begin{algorithmic}
	\If{$\sum\limits_{i = 1}^{N} d_i \leq B$}
	\State  MEC confirms $d_i$ as $d_i^{*}$
	\Else 
	\State  MEC sets $d_i^{*} = d_i \frac{B}{\sum\limits_{i = 1}^{N} d_i}$ 
	\EndIf
	\For{$k = 1, 2, 3, ...$ }
	\For{each $i \in \{1,2,3,...,N\}$}
	\State \textbf{Get} $U_j'(x_j(k))$ from all neighbours, $j \in {N_k^i}$
	\State \textbf{Calculate} $q_i(k) = \eta _i \sum\limits_{j \in {N_k^i}} (U_j'(x_j(k)) - U_i'(x_i(k)))$
	\State \textbf{Calculate} $\Delta x_i(k) = x_i(k) - d_i^{*}$
	\State \textbf{Update} $ U_i'(x_i(k+1)) =  U_i'(x_i(k)) + q_i(k) - \zeta_i(k) - \mu \Delta x_i(k)$
	\State \textbf{Update} $\zeta_i(k+1) = \zeta_i(k) - \mu q_i(k)$ 
	\State \textbf{Calculate} $x_i(k+1)$ using $U_i'(x_i(k+1))$
	\EndFor
	\EndFor
\end{algorithmic}
\label{alg1}
\end{algorithm}

The algorithm starts by determining whether the overall requested $d_i$ exceeds the total available bandwidth $B$. If so, the MEC server repartitions each request proportionally. This step is to ensure that the constraint in Equation \eqref{eq} is satisfied. Upon receiving the updated $d_i^{*}$, each IoT device exchanges its derivative information, $U'_i(x_i(k))$, with its neighbours, denoted by $N_k^{i}$, until the algorithm converges to optimality. This step is fully distributed as each device only relies on its own information and that of its neighbours. In other words, there is no continuous feedback required by the MEC server once the $d_i^{*}$ is confirmed. This operation significantly reduces  communication overheads and improves robustness of the overall system.

\section{Simulation}

\begin{figure}[htbp]
	\centering
	\vspace{-0.1in}
	\hspace{-0.3in}
	\subfigure{
		\begin{minipage}[t]{0.52\linewidth}
			\centering
			\includegraphics[width=2in]{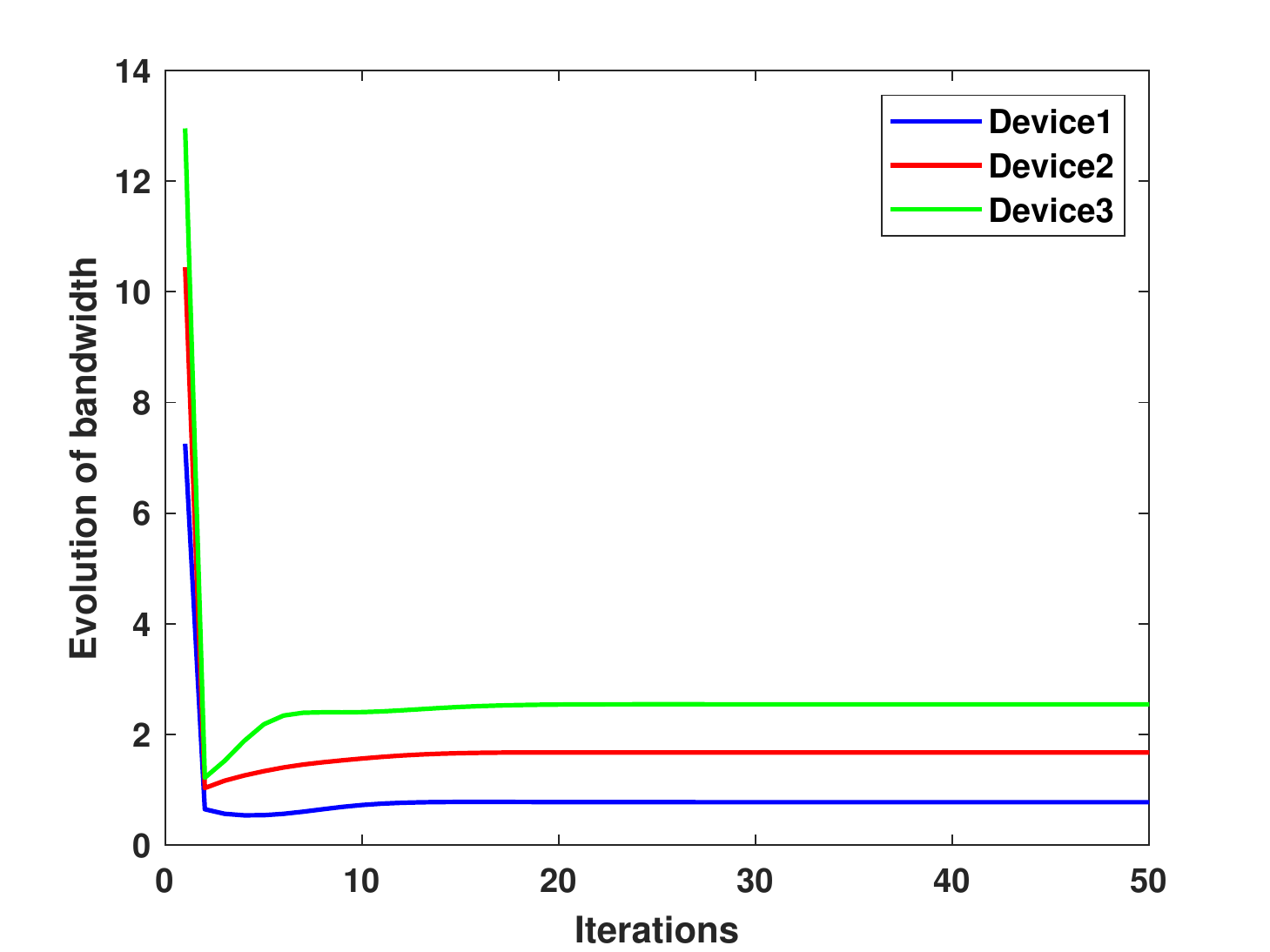}
		\end{minipage}%
	}%
	\subfigure{
		\begin{minipage}[t]{0.52\linewidth}
			\centering
			\includegraphics[width=2in]{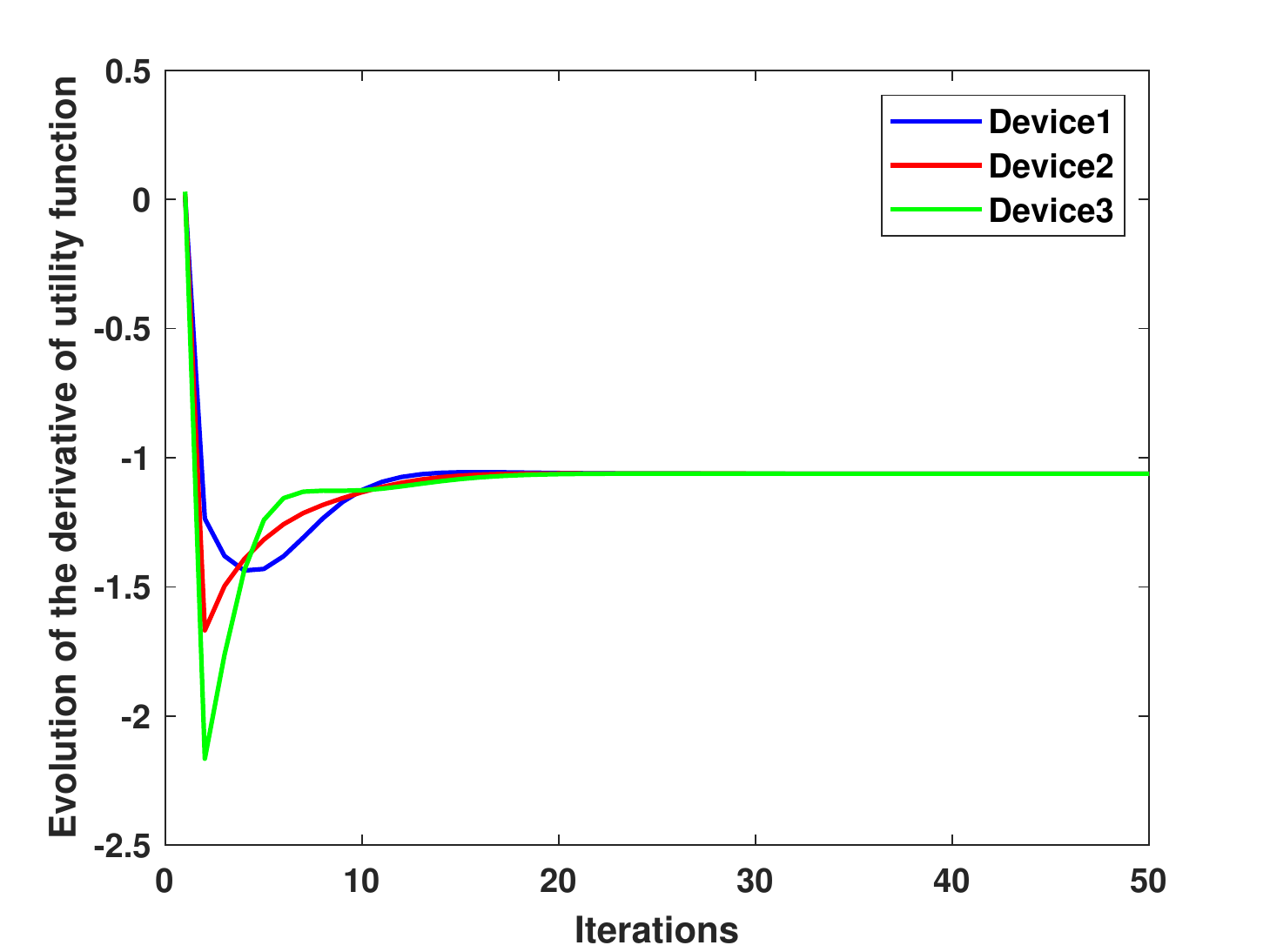}
		\end{minipage}%
	}%

	\centering
	\caption{Evolution of the optimization process for (left) bandwidth and (right) derivative of the utility function.}
	\vspace{-0.2in}
	\label{fig_evolution}
\end{figure}

This section presents simulation results to evaluate the performance of the proposed system. The algorithm is implemented using Matlab for a simple scenario using only 3 IoT devices as a proof of concept. We assume that all devices can have bidirectional communication links with their neighbours, where device 1 can connect to device 2, device 2 can communicate to both 1 and 3, and device 3 can only connect to device 2. The system parameters are set as follows: $N=3$, $B = 5$, $\frac{S}{R} = 100$, $\omega_1 = 1$, $\omega_2 = 2$, $\omega_3 = 3$, $d_1 = 1$ , $d_2 = 2$, $d_3 = 2$, $p = 0.01$, $\mu = 0.2$. Our optimization result is shown in Fig. \ref{fig_evolution}, indicating that the proposed algorithm can rapidly converge to the theoretical optimal result with $x_1^* = 0.78$, $x_2^* = 1.67$, $x_3^* = 2.55$. The consensus derivative value shown on the right of the plot depicts the optimality of our converged result (KKT condition).

\section{Conclusion}\label{sec:Conclusion}

In this paper, we propose a novel bandwidth allocation system for IoT edge devices. The system is able to automatically find the optimal bandwidth for each IoT device in a resource-constrained NB-IoT network. Different algorithms and complex pricing models will be further investigated as part of our future work. 

\section*{Acknowledgement}

This work was support in part by SFI Grant SFI/12/RC/2289\_P2 and Hongde Wu is supported by the DCU research master scholarship.

\bibliographystyle{ieeetr}
\bibliography{References}

\end{document}